\documentstyle{laa}
\thesaurus{12.03.4; 11.06.1}
\title{Substructure effects on the collapse of density perturbations\\ } 
\author{A. ~Del Popolo\inst{}, M. Gambera\inst{} }
\offprints{A. ~Del Popolo, M. ~Gambera}
\institute{Istituto di Astronomia dell'Universit\`a di Catania, \\
Citt\`a Universitaria, Viale A.Doria, 6 - I 95125 Catania, Italy}
\date{}
\begin{document}
\maketitle
\begin{abstract}
We solve numerically the equations of motion for the collapse 
of a shell of baryonic matter, made of galaxies and substructure 
of $ 10^{6} M_{\odot}- 10^{9} M_{\odot}$, falling into 
the central regions of a cluster of galaxies  
taking account of dynamical friction.
We calculate the evolution of the expansion parameter, a(t),  
of the perturbation using a coefficient of dynamical friction, $ \eta_{0}$, 
calculated for a perturbation in which 
clustering is absent and a coefficient $ \eta_{cl}$ obtained 
from a clustered one. 
The effect of the dynamical friction is to 
slow down the collapse (V. Antonuccio-Delogu \& S. Colafrancesco 1994, 
hereafter AC) 
producing an observable variation of the parameter of expansion 
of the shell. The effect increases with increasing $ \eta $ and 
with the increasing of clustering. Finally, we show how the collapse 
time depends on $ \eta_{0}$ and $ \eta_{cl}$.\\      
\keywords{cosmology: theory- 
galaxies: formation}
\end{abstract}
\begin{flushleft}

{\bf 1.Introduction \\}

\end{flushleft}

\vspace*{0.5 mm}

\noindent
In the most promising cosmological scenarios, structure formation 
in Universe is generated through the growth and collapse of primeval density 
perturbations originated from quantum fluctuations (Guth \& Pi 1982; 
Hawking 1982; Starobinsky 1982; Bardeen, Bond, Kaiser and Szalay 1986, hereafter
BBKS) in an inflationary phase of early Universe. 
Density fluctuations originated in the inflationary era are Gaussian 
distributed and their statistics  
can be expressed entirely 
in terms of the power spectrum of density fluctuations: 
\begin{equation}
P( k) = < |\delta_{{\bf k}}|^{2}> 
\end{equation}
where 
\begin{equation}
\delta_{{\bf k}} =\int d^{3} k exp(-i {\bf k x}) \delta({\bf x})
\end{equation}
\begin{equation}
\delta({\bf x}) = \frac{ \rho ({\bf x}) - \rho_{b}}{ \rho_{b} }
\end{equation}
and $ \rho_{b} $ is the mean background density. \\
In biased structure formation theory it is assumed that cosmic structures 
of linear scale $ R$ form around the peak of density field, 
$  \delta( {\bf x})$, smoothed on the same scale. 
Density perturbations evolve towards non-linear regime because of 
gravitational instability which breaks away from general expansion 
at a time, $ t_{m}$, given by: 
\begin{equation}
t_{m} = \left[\frac{ 3 \pi}{32 G \rho_{b}} ( 1 +\overline{\delta})
\right]^{1/2} (1+z)^{3/2}
\end{equation}
where $ z$ is the redshift, $ \overline{\delta}$ is the overdensity 
within $ r$. When $ \overline{\delta} \simeq 1$ the density 
perturbation begins to collapse. The collapse time  
depends on the characteristic of initial fluctuation field  such as 
the average overdensity and on the environment in which the perturbation 
is embedded. This last feature depends on the cosmological scenario. \\
In this paper, we consider the CDM model (Liddle \& Lyth 1993) based on a scale 
invariant spectrum of density fluctuations growing under 
gravitational instability. 
The modality of the collapse of a perturbation depends on the chosen 
model. 
A very simple model for 
accretion of matter in cluster of galaxies was firstly investigated 
by Gunn \& Gott (1972). 
Gunn \& Gott's model 
is an oversimplification of the perturbation evolution.   
In their model they considered only the collapse of spherical uniform overdense
shells of matter 
surrounding an isolated top-hat perturbation embedded within a 
homogeneous background without including the tidal 
interactions among the shell of matter 
and the external density perturbations and 
excluding
the presence of substructure 
(collapsed objects whose length is shorter than that of the  
main perturbation). This does not take into account 
the previsions 
of CDM models according to which there is 
an abundant 
production 
of substructure during the evolution of the Universe.\\ 
In this scenario structure formation proceeds  
bottom-up through gravitational clustering, merging and violent 
relaxation of small scale substructure (White \& Rees 1978). 
So the matter 
inside a given region is clumped in a hierarchy of objects of various 
dimensions.
The substructure acts as a source of stochastic fluctuations 
in the gravitational field of the protostructure 
inducing dynamical friction (AC) and eventually producing a modification 
of the motion of shells of matter in a density perturbation. 
The galaxies inside a shell of matter are subject to the 
stochastic gravitational field produced by the substructure and 
their motion undergoes a preferential deceleration in the direction 
of motion. 
In particular, dynamical friction produces a delaying effect 
in the collapse of the regions with $ \overline{\delta}
\leq 10^{-2}$ inside the perturbation as shown in (AC). The final result is that Gunn \& 
Gott's model is inadequate to describe the collapse of a spherical 
perturbation and it requires a revision. \\
To this aim we used the modified equation of motion   
of a shell of baryonic matter given by AC, 
in which a frictional force that takes into 
account dynamical friction effects is introduced
(we study only the component of the dynamical friction due to the galaxies
belonging to the shell and to the substructure, neglecting gravitational bound to the
shell) and we solve it by numerical integration. \\
In this paper we show: \\  
a) how the expansion parameter, $ a(t)$, 
of the shell is changed by the presence of substructure both in 
the case of an unclustered and in a clustered system; 
b) how clustering increases the effects of dynamical 
friction;
c) the changes produced 
by dynamical friction on  
the collapse time of the perturbation and the role 
of clustering in this process. \\ 
\begin{flushleft}
{\bf 2. Modification of Gunn \& Gott's equation.\\}
\end{flushleft}
In a hierarchical structure formation model, the large 
scale cosmic environment can be represented as a collisionless medium 
made of a hierarchy of density fluctuations 
whose mass, $ M$, is given by the mass 
function $ N(M,z)$, where z is the redshift. In these models 
matter is concentrated in lumps, and the lumps into groups and so on. 
In what follows, we consider a shell of matter outside 
the main body of a perturbation which collapsed 
to form a cluster of galaxies.\\ 
The equation of motion of a shell of matter (as previously 
said composed of galaxies and substructure) around a 
maximum of the density field (neglecting tidal interactions 
and the substructure), can be expressed in the form:
\begin{equation}
\frac{d^{2} r}{d t^{2} }= -\frac{G M}{r^{2}(t)} \label{eq:pee}
\end{equation}
(Peebles 1980, Eq. ~19.9), where $ M$ is the mass enclosed with  the proper 
radius $ r(t)$. Using Gunn \& Gott's notation the proper radius 
can be written as: 
\begin{equation}
r(r_{i},t) = 
a( r_{i}, t) r_{i} 
\end{equation}
where $ r_{i} $ is the initial radius and $ a(r_{i},t)$ is the expansion  
parameter of the shell. At the initial time $ t_{i}$ the initial 
condition is given by 
\begin{equation}
a(r_{i},t_{i}) =1
\end{equation}
In the presence of substructure it is necessary 
to modify the equation of motion, Eq. ~(\ref{eq:pee}), because the graininess 
of mass distribution in the system induces dynamical 
friction that at last introduces 
a frictional force term in Eq. ~(\ref{eq:pee}).\\
In a material 
system, the gravitational field can be decomposed into an average field, 
$ {\bf F}_{0}(r)$, generated from the smoothed out distribution 
of mass, and a stochastic component, $ {\bf F}_{stoch}(r)$, generated 
from the fluctuations in number of the field 
particles. 
The stochastic component of the gravitational field is 
specified assigning a probability density, $ W( {\bf F})$, (Chandrasekhar \& 
von Neumann 1942). In an infinite homogeneous unclustered system 
$ W( {\bf F})$ has been given by Holtsmark distribution 
(Chandrasekhar \& 
von Neumann 1942) while in inhomogeneous and clustered systems $ W({\bf F})$ 
has been given by Kandrup (1980) and Antonuccio-Delogu \& Barandela (1992) respectively. 
The stochastic force, $ {\bf F}_{stoch}$, in a self-gravitating 
system modifies the motion of particles as it is done by a frictional force. 
In fact a particle moving faster than its neighbours produces a deflection 
of their orbits in such a way that the average density is greater in 
the direction opposite that of traveling 
causing a slowing down 
in its motion. \\
Following  Kandrup (1980) in the hypothesis 
that there are no correlations 
among random force and its derivatives the frictional force is given by:
\begin{equation}
{\bf F} = -\eta {\bf v} = - \frac{ \int W(F) F^{2} T(F) d^{3} F}{2 < v^{2}>}
{\bf v}  
\end{equation}
where $ \eta$ is the coefficient of dynamical friction, 
$ T(F)$ the average duration of a random force impulse, $ <v^{2}>$ 
the characteristic speed of a field particle having a distance 
$ r \simeq (\frac{G M}{F})^{1/2}$ from a test particle (galaxy). 
This formula permits to calculate 
the frictional force
for inhomogeneous systems when 
$ W(F)$ is given. If the field particles are distributed homogeneously 
the dynamical friction force is given by: 
\begin{equation}
F = -\eta  v = -\frac{ 4.44 G^{2} m_{a}^{2} n_{a}}{[<v^{2}>]^{3/2}} 
\log \left\{ 1.12 \frac{<v^{2}>}{ G m_{a} n_{a}^{1/3}}\right\}
\end{equation}
(Kandrup 1980), 
where $ m_{a}$ and $ n_{a}$ are respectively the average mass and density of 
the field particles. Using the virial theorem we 
can written
the dynamical friction force  as follows: 
\begin{equation}
F = -\eta  v = -\frac{ 4.44 [G m_{a} n a^3]^{1/2}}{N} 
\log \left\{ 1.12 N^{2/3} \right\} \frac{v}{ a^{3/2}}
\end{equation}
where $ N$ is the total number of field particles.    
In this last equation it is assumed that the field particles generating 
the stochastic field are virialized. This is justified by  
the previrialization hypothesis (Davis \& Peebles 1977). \\
To calculate the dynamical evolution of the galactic component 
of the cluster it is necessary to calculate the number and average
mass of the field particles generating the stochastic field. \\
The protocluster, before the ultimate collapse at $ z \simeq 0.02$, is made 
of substructure having masses ranging from $ 10^{6} - 10^{9} M_{\odot}$
and of galaxies. We suppose that the stochastic gravitational 
field is generated from that portion of substructure having a 
central height $ \nu$ larger than a critical threshold $ \nu_{c}$. This 
latter quantity can be calculated (following AC) using the condition that the 
peak radius, $ r_{pk} (\nu \ge \nu_{c}), $ is much less than the average 
peak separation $ l_{av} \equiv n_{a} (\nu \ge \nu_{c})^{-1/3}$
(see AC and BBKS).
The condition $ r_{pk} (\nu \ge \nu_{c}) < 0.1n_{a}(\nu \ge \nu_{c})^{-1/3}$ 
ensures that the peaks of substructure are point like. Using 
the radius for a peak (AC Eq. ~13) that is: 
\begin{equation}
r_{pk} = \sqrt{2} R_{\ast} \left[ \frac{1}{(1+\nu \sigma_{0}) 
(\gamma^{3} +(0.9/\nu))^{3/2}}\right]^{1/3}
\end{equation}
where $ \gamma$, $ R_{\ast}$ are parameters related to moments of the 
power spectrum (BBKS Eq. ~4.6A),
we obtain a value of $ \nu_{c} =1.3$ and then we have 
$ n_{a} (\nu \ge \nu_{c} ) =50.7 Mpc^{-3} $ 
($ \gamma =0.4$, $ R_{\ast}=50 Kpc$) and $ m_{a}$ is given by:
\begin{equation}
m_{a} = \frac{1}{n_{a} (\nu \ge \nu_{c})} \int_{\nu_{c}}^{\infty}  
m_{pk}(\nu)N_{pk}(\nu)d\nu = 10^{9} M_{\odot}
\end{equation}
(in accordance with the result of AC), where $ m_{pk}$ is given in Peacock 
$ \&$ Heavens (1990) and $ N_{pk}$ is the average number density of peak
(BBKS Eq. ~4.4).
Now we study the evolution of a shell of matter, made of 
substructure and galaxies, as modified by dynamical friction. \\
As previously told, we suppose that the substructure has a Maxwellian 
distribution of velocity and that the galaxy moves in the substructure background.
The modified equation of motion for each galaxy of a shell of matter 
can be 
written in the form:
\begin{equation}
\frac{d^{2} { r}}{d t^{2} }= -\frac{G M}{r^{2}(t)} -\eta v \label{eq:pees}
\end{equation}
(Langevin 1908, Kandrup 1980, Saslaw 1985, Kashlinsky 1986, 
AC) 
In the spherical collapse model, it is supposed that the infall is radial 
for every galaxy in the protocluster. Eq. (~ \ref{eq:pees}) describes 
the motion of each  galaxies, the only parameter that 
describes the collapse is the radial distance, $ r$, or the parameter 
$a(t)$ previously defined.  
Remembering that the average density is given by:
\begin{equation}
\overline{\rho}(r_{i}, t) = \frac{3 M}{4 \pi a^{3}(r_{i},t) r_{i}^{3} } 
\end{equation}
and that mass conservation requires:
\begin{equation}
\overline{\rho} (r_{i}, t_{i})= 
\frac{\overline{\rho} (r_{i}, t_{i})}{a^{3}(r_{i}, t)}
\end{equation}
equation 
( \ref{eq:pees}) in terms of $ a( r_{i} ,t)$  
can be written as:
\begin{equation}
\frac{d^{2} a}{d t^{2} }= -
\frac{4 \pi G \rho_{ci}( 1+\overline{\delta_{i}})}{3 a^{2}(t)} -
\eta \frac{ d a }{ d t} \label{eq:peesl}
\end{equation}
where $ \rho_{ci} $ is the background density at a time $ t_{i} $ 
and $ \overline{ \delta}_{i}$ is the overdensity within $ r_{i}$. 
Using the parameter $ \tau = \frac{t}{T_{c0}}$, 
where $ T_{c0}/2$ is the collapse time in absence of dynamical friction
(Gunn \& Gott Eq. 16),
Eq. ~(\ref{eq:peesl}) 
may be written in the form: 
\begin{equation}
\frac{d^{2} a}{d \tau^{2} }= -
\frac{4 \pi G \rho_{ci}( 1+\overline{\delta_{i}})}{a^{2}(\tau)} T_{c0}^{2}-
\eta_{0} T_{c0} \frac{ d a }{ d \tau} ( 1+ \frac{\eta_{cl}}{\eta_{0}}) \label{eq:pessr}
\end{equation}
Equation (\ref{eq:pessr}) is obtained remembering that the probability 
density $W(F)$ depends linearly on the correlation function 
(AC) and so it is 
possible to decompose the coefficient of dynamical friction, $ \eta$, 
as follows:
\begin{equation}
\eta = \eta_{0} + \eta_{cl} 
\end{equation}
where $ \eta_{0}$ is the coefficient 
of dynamical friction of an unclustered distribution of field particles, 
while $ \eta_{cl}$ takes into account clustering. 
\begin{flushleft}
{\bf3. Results and discussions.\\}
\end{flushleft}
The time evolution of the expansion parameter, $ a(\tau) $, can be 
obtained solving Eq. ~(\ref{eq:pessr}). \\
There are two ways to solve 
equation (\ref{eq:pessr}). The first one is to  
look for an asymptotic expansion, the second one is numerical.    
In AC the authors 
gave only an expression for the collapse time, $ T_{c}$,
taking into account dynamical friction. We solved Eq. ~(\ref{eq:pessr}) 
numerically using 
a Runge-Kutta integrator 
of $ 4^{th}$ order and we studied the motion of a shell of matter 
of low density, 
$ \overline{\delta} =0.01$, typical of a perturbation present 
in the outskirts of a cluster of galaxies. We chose the 
initial conditions 
remembering that 
at the maximum of expansion the initial 
velocity is zero. 
We show in fig. ~1  the results of our calculation 
when the dynamical friction is present but considering 
only the case of an unclustered distribution ( $ \eta_{0} \neq  0$ and 
$ \eta_{cl} = 0 $). \\
In this case the expansion parameter of a shell, 
$a( \tau) $,  
is plotted 
versus $ \tau$, for different values of $ \eta_{0}$, the coefficient of 
dynamical friction for an unclustered system (stochastic force 
generators randomly distributed).
\begin{figure}
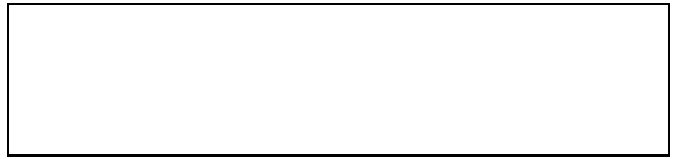

\picplace{2.0cm}
\caption[]{Temporal evolution of the expansion parameter 
of a shell of matter made of galaxies and 
unclustered substructure. The solid line is $ a(\tau)$ when 
dynamical friction is absent while the dotted line is $ a(\tau)$
when it is taken into account ( $ \eta_{0} \neq  0$ and 
$ \eta_{cl} = 0 $).
The dashed line shows the effect of dynamical friction when 
 $ \eta'_{0} >  \eta_{0} $. 
We 
assume a cluster radius of $R_{cl} = 5 h^{-1} Mpc $, a central overdensity 
$\overline{ \delta} =0.01$ and 
a total number of peaks of substructure $ N_{tot} = 10^{3}$. }
\end{figure}

The effect of dynamical friction 
is to make the decrease of $ a(\tau)$ less steep and consequently 
to slow down the collapse of the shell in agreement with the analytic 
calculation performed by AC. 
The evolution 
of the parameter of expansion, $ a(\tau)$, is more 
and more modified as $ \eta_{0}$ increases, in comparison
to Gunn \& Gott's model, 
as one can see from figure 1.\\ 
In Fig. ~2 we show the same parameter, 
$a( \tau)$, for different values of $ \eta_{cl}$, the coefficient 
of dynamical friction of a clustered system (clustering of substructure 
is required, for example, by the minihalo model (Mo et al. 1993)).    
\begin{figure}
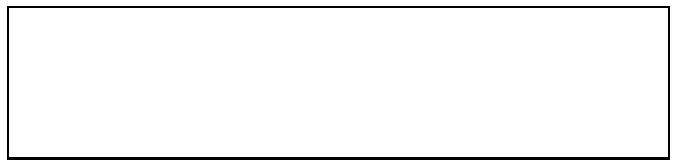

\picplace{2.0cm}
\caption[]{Temporal evolution of the expansion parameter of a shell 
of matter made of galaxies and clustered substructure. 
The solid line is $ a(\tau) $ when dynamical 
friction is absent while the dotted line is the same when it is taken 
into account ($ \eta_{cl} \neq 0 $).
The dashed line shows the effect of dynamical friction when 
$ \eta'_{cl} >  \eta_{cl} $.
We 
assume a cluster radius of $R_{cl} = 5 h^{-1} Mpc $, a central overdensity 
$\overline{ \delta} =0.01$,  and 
a total number of peaks of substructure $ N_{tot} = 10^{3}$}.
\end{figure}
Clustering produces a sensible slowing down of the shell
collapse with respect to unclustered systems ($ \eta_{cl}= 0$). 
This result is perfectly understandable since the introduction 
of positive clustering produces a greater probability for a test particle 
to be scattered during its motion, increasing the role 
of dynamical friction. \\ 
The caracteristic values for $ \eta$ are between $0 < \eta < 3$, they
depend strongly on the value of many parameters like  $ \nu_c$ and
others (Gambera et al. 1996, in preparation).  
For a typical protocluster configuration  like that of 
Coma, 
that can be considered almost as a lower limit
having an initial overdensity, at $ z \simeq 10^{3}$, given by 
$ \overline{\delta} \simeq 8 \cdot 10^{-3}$   and a total mass estimated 
in $ M \simeq 10^{15} M_{\odot}$ (Hughes 1989), supposing that 
small scale substructure 
is made of objects of $ M \simeq 10^{9} M_{\odot}$,
we obtain $ \eta ~\simeq 0.2$.\\
Finally, we computed the time of collapse of a shell, $T_{c}$,  
versus both 
$ \eta_{0} $ and of $ \eta_{cl} $. 
To this aim we solved Eq. ~(\ref{eq:pessr}) varying $ \eta_{0} $ 
and for each value of this parameter we obtained the zero of $ a(\tau)$, 
that gives us $T_{c}$. 
We  repeated the same procedure for $ \eta_{cl}$.  
In Fig. ~3 we show the results of the computation. \\ 
It is interesting to observe that $ T_{c}$ is very sensitive to the changes
of the coefficent of dynamical friction for a clustered system $ \eta_{cl} $
and increases with increasing $ \eta_{cl} $ while it is less sensitive to
the changes of the coefficent of dynamical friction for an unclustered system.
Dynamical friction changes the dynamical evolution 
of low density perturbations ($\overline{\delta} \simeq 0.01$) and 
clustering contributes to increase this effect. 
\begin{figure}
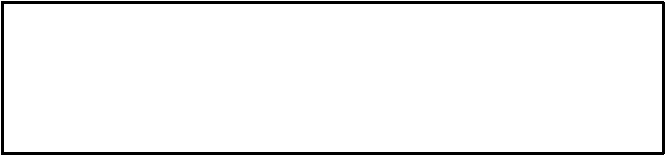

\picplace{2.0cm}
\caption[]{The variation of time of collapse $ \tau = \frac{t}{T_{c0}}$
 with $ \eta$. The solid line is the time of collapse of an unclustered 
system in which dynamical friction is present while the dotted line is
the same for a clustered system and the dashed line is the time 
of collapse for a system in which dynamical friction is absent}
\end{figure}
~\\
\begin{flushleft}
{\bf 4. Conclusions} \\
\end{flushleft}
In this paper we have discussed the dynamics of the infall of 
a shell of baryonic matter of a spherical perturbation, 
when the effects of dynamical friction are taken into account, 
using the time evolution of the parameter of expansion, $ a(\tau) $.  
In particular,
we showed that the parameter  $a(\tau)$  
of a shell of matter of a protocluster decreases less steeply with increasing
$ \eta_{0}$ and that clustering produces a further increase in this effect. 
This effect, then, could be the cause because many clusters of
galaxies are not yet relaxed.
Finally,  
we showed how the collapse time, $ T_{c}$,  varies in presence 
or absence of dynamical friction (both in the case in which 
$ \eta_{cl}= 0$ and when 
$ \eta_{cl}\neq 0$). \\
The collapse time, $ T_{c}$, of an infalling shell increases with 
increasing values of $ \eta_{0}$ (see Fig. 3). Clustering enlarges 
the change in the dynamical evolution of the perturbation 
produced by dynamical friction. An interesting point on which we
are in progress (Gambera et al. 1996, in preparation) is the
determination of how
the growth of the collapse time depends 
on $ \nu_{c}$ and on others parameters. Besides, we want to 
find an analytic relation that links $ \tau$ with the 
parameters on which it dependes. \\

\begin{flushleft}
{\it Acknowledgements}.
We want to thank V. Antonuccio-Delogu for his helpful and stimulating
discussions during the period in which this work was performed
and the referee Dr. P. S\'eguin for several useful comments.\\
\end{flushleft}

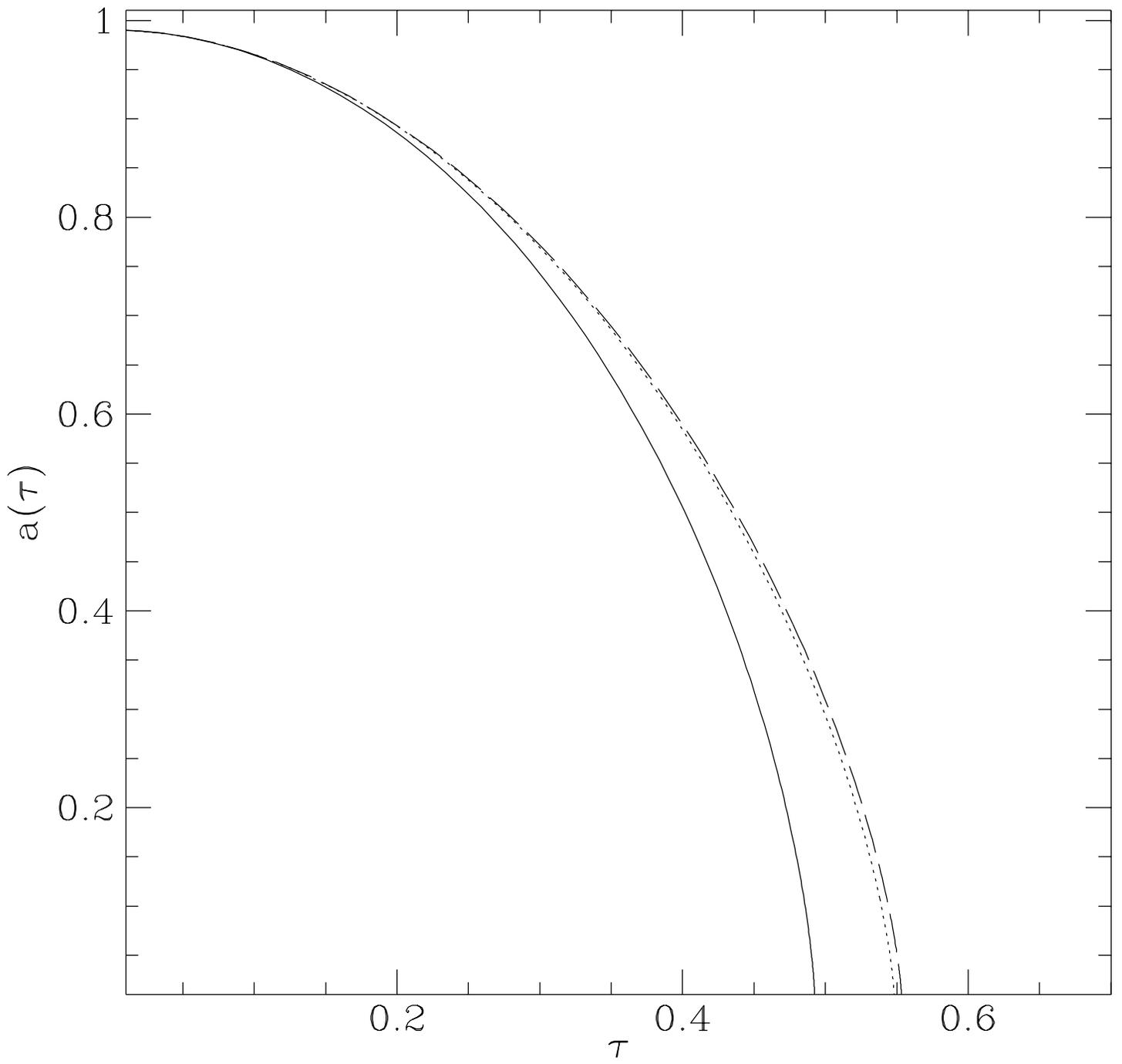
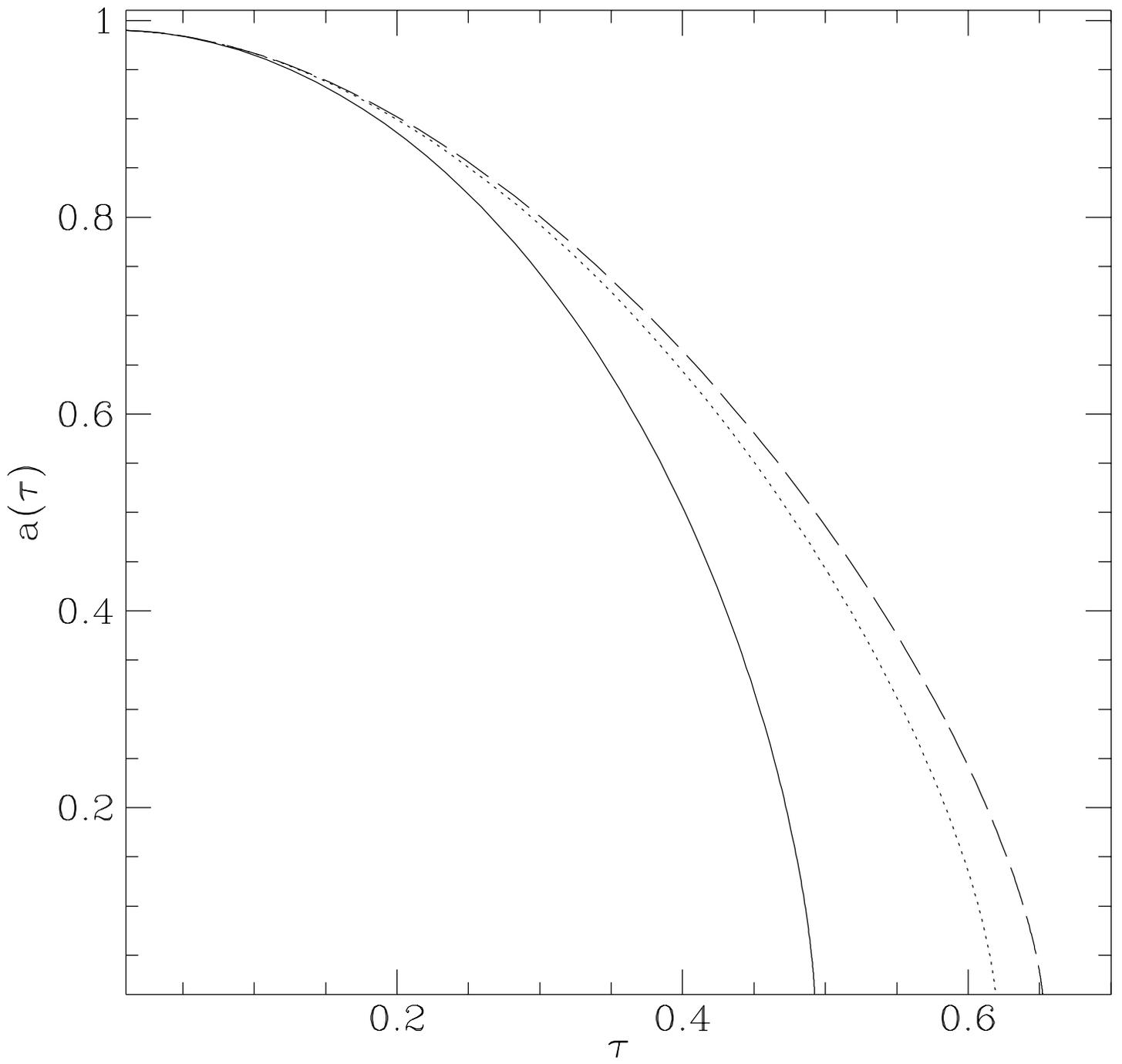
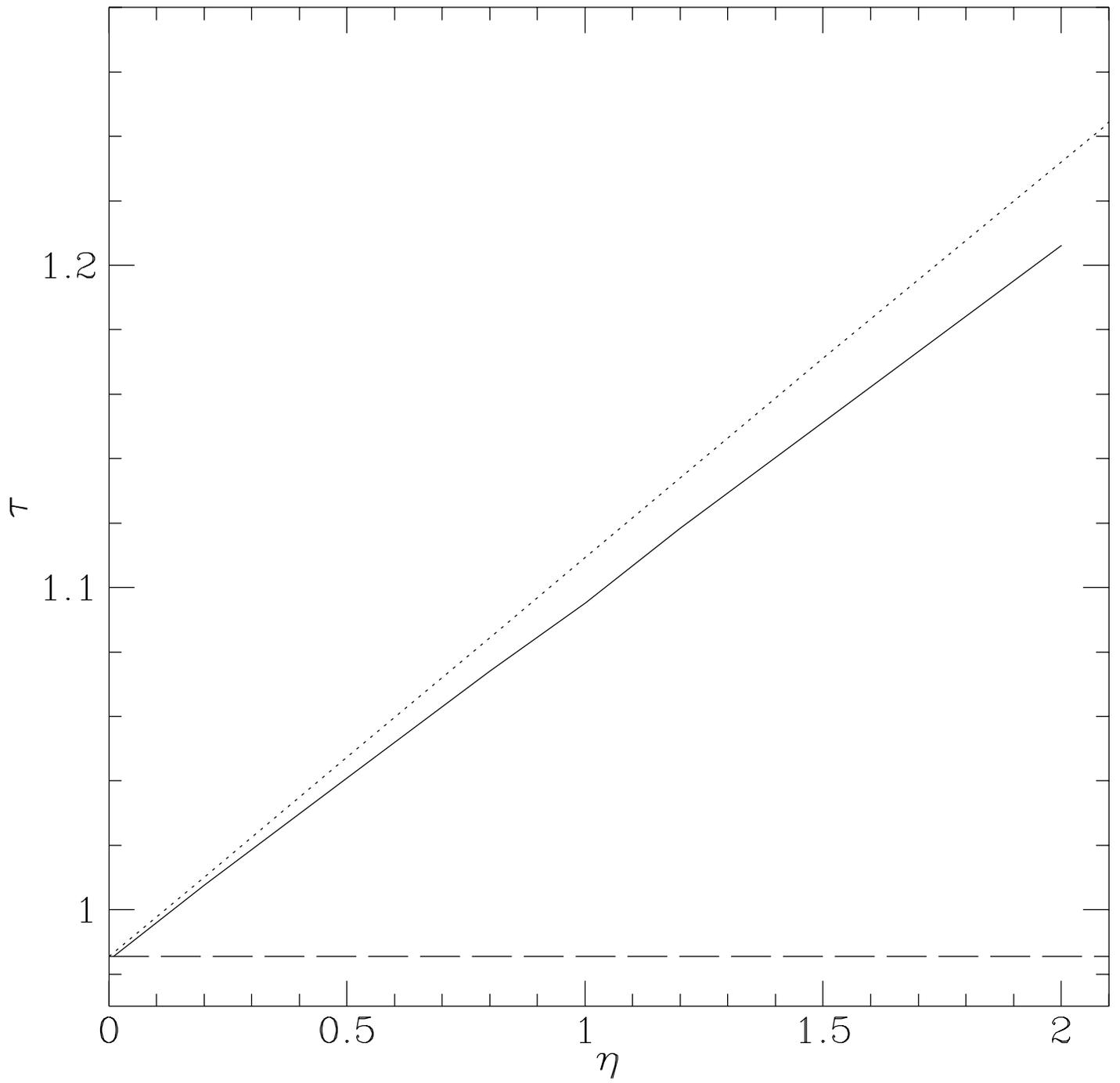
\end{document}